\documentclass[11pt]{article}
\usepackage{sao1}

\usepackage{graphicx}

\begin{document}

\title{Investigating the pre-main sequence magnetic chemically peculiar system HD 72106}
\author{Folsom, C. P.\inst{1,2} \and Wade, G. A.\inst{1} \and Hanes, D. A.\inst{2} \and 
  Catala, C.\inst{3} \and  Alecian, E.\inst{3} \and Bagnulo, S.\inst{4} \and Boehm, T.\inst{5} \and  Bouret, J.-C.\inst{6} \and Donati, J.-F.\inst{5} \and Landstreet, J. D.\inst{7}	}
\institute{ Dept. of Physics, Royal Military College of Canada, PO Box 17000, Stn Forces, Kingston, Ontario, Canada, K7K 7B4
	    \and Dept. of Physics, Engineering Physics \& Astronomy, Queen's University, Kingston, Ontario, Canada, K7L 3N6
	    \and Observatoire de Paris, LESIA, 5 place Jules Janssen, 92190 Meudon CEDEX
	    \and European Southern Observatory, Casilla 19001, Santiago 19, Chile
	    \and Observatoire de Midi-Pyr\'en\'es, 14 Av. E. Belin, F-31400 Toulouse, France 
	    \and Laboratoire d'Astrophysique de Marseille, Traverse du Siphon, BP 8, 13376 Marseille Cedex 12, France
            \and Physics \& Astronomy Department, University of Western Ontario, London, Ontario, Canada, N6A 3K7 }
\maketitle

\begin{abstract}
The origin of the strong magnetic fields observed in chemically peculiar 
Ap and Bp stars stars has long been debated.  The recent discovery of 
magnetic fields in the intermediate mass pre-main sequence Herbig Ae and Be 
stars links them to Ap and Bp stars, providing vital clues about Ap and Bp stars and the origin and evolution
of magnetic fields in intermediate and high mass stars.  A detailed study of one young magnetic B star, HD~72106A, 
is presented.  This star appears to be in a binary system with an apparently normal Herbig Ae star.  
A maximum longitudinal magnetic field strength of $+391 \pm 65$ G is found in HD~72106A, 
as are strong chemical peculiarities, with photospheric abundances of some elements ranging up to 100x above solar.  
\end{abstract}

\section{Introduction: HAeBe and Ap stars}

A few percent of main sequence A and B stars display strong, globally ordered 
magnetic fields; these are the so-called Ap and Bp stars.  The 
origin and evolution of these magnetic fields is currently the subject 
of intensive research.  An important avenue of this research 
is the investigation the progenitors of Ap and Bp stars.
Herbig Ae and Be (HAeBe) stars are the pre-main sequence progenitors 
of the main sequence A and B type stars.  As such, HAeBe 
stars have intermediate masses and display emission, infrared excess, 
and tend to be associated with dust and nebulosity.  It has been long suggested 
that some HAeBe stars may evolve into magnetic main sequence Ap and Bp 
stars.  If this were the case, it was hoped that there would be some 
distinguishing observable feature in HAeBe stars linking them to Ap stars. 

Pioneering observations in 2005 and 2006 successfully detected 
magnetic fields in HAeBe stars for the first time  
(Wade et al. 2005, Catala et al. 2006).  The detected fields appear to display 
similar intensities and geometries when compared to Ap stars 
(Alecian et al. 2006).  The longitudinal field strengths 
detected are on the order of hundreds of gauss, and the geometries 
are predominantly dipolar.  Additionally, the incidence of magnetic 
fields in HAeBe stars is similar to that 
of the magnetic Ap and Bp stars.  
This strongly supports the idea that magnetic HAeBe stars do in fact 
represent the progenitors of main sequence magnetic Ap and Bp stars. 

In this paper we investigate one particularly interesting young magnetic B star, HD~72106A, in detail.  

\section{HD 72106}

\begin{figure}[!htb]
\centering
\includegraphics[width=4.0in]{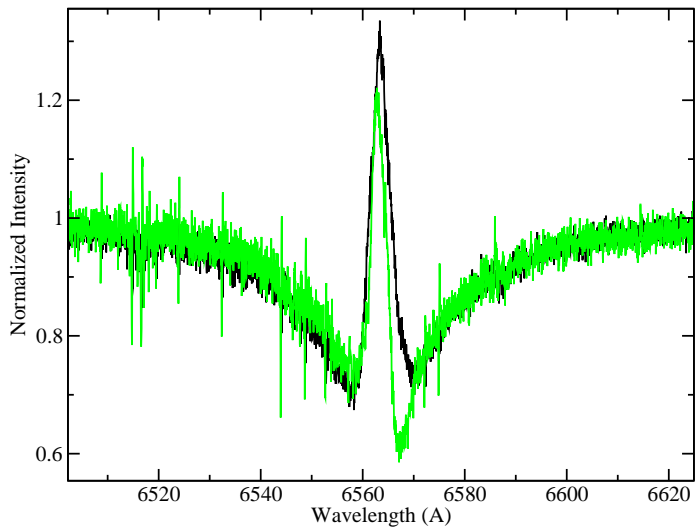}
\caption{Observed H$\alpha$ profiles for HD 72106B (the apparently non-magnetic secondary) from 11 Jan. 2006 (dark/black) and 11 Feb. 2006 (light/green).
Strong and variable emission is observed, indicating the presence of an emitting circumstellar envelope.  }
\label{h-alpha}
\end{figure}

HD~72106 is a visual double system, with a 0.8'' separation.  The primary component has a definitely 
detected magnetic field (Wade et al. 2005), and the secondary is clearly a Herbig Ae/Be (HAeBe) 
star (Vieira et al. 2003). The secondary shows strong, variable emission in H$\alpha$, as shown in Fig. \ref{h-alpha}.  

The system has a very large proper motion in $\alpha$: $-5.18 \pm 1.08$ mas/yr, 
and in $\delta$: $9.73 \pm 1.36$ mas\,y$^{-1}$ 
but no relative motion between components is observed (Hartkopf et. al. 1996 \& Hipparcos data, H\o g et al. 2000).  The radial 
velocities of the both components are identical to within $\pm 1$ km\,s$^{-1}$, at 22 km\,s$^{-1}$.  
As well, the Hipparcos parallax solution for the system places the two 
stars at the same distance: $288 (+202/-84)$ pc.  
This indicates that the two stars are travelling in tandem, and strongly suggests that the system is in fact a binary.  This is critical because, 
if the system is truly binary, then both components likely have the same age and initial formation conditions.  

Wade et al. (2006) determine, for the primary, an effective temperature of $11000 \pm 1000$ K, 
a $\log\,{g}$ of $4.0 \pm 0.5$, and a mass of $2.4 \pm 0.4$ $M_\odot$.  
For the secondary they derive an effective temperature of $8000 \pm 500$ K, 
a $\log\,{g}$ of $4.25 \pm 0.25$, and a mass of $1.75 \pm 0.25$ $M_\odot$. 

\begin{figure}[!htb]
\centering
\includegraphics[width=4.5in]{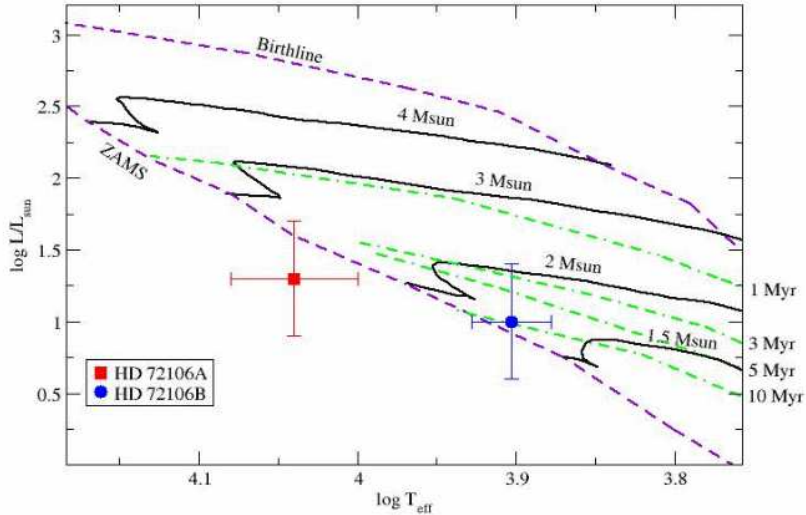}
\caption{The pre-main sequence H-R diagram of the HD~72106 system.  The solid lines are the model 
evolutionary tracks of Palla and Stahler (1993), with masses as labeled, and 
the dot-dashed lines are isochrones, with ages labeled.  The birthline (which
assumes an accretion rate of $10^{-5} M_\odot\, {\rm yr}^{-1}$) and zero age main sequence (ZAMS) 
are shown as dashed lines.  }
\label{h-r}
\end{figure}

The two stars can be placed on an H-R diagram 
(Wade et al. 2006), and have consistent 
ages of about 10 Myr, near the zero-age main 
sequence, when compared with isochrones from Palla \& Stahler (1993).  
This is consistent with the components being co-eval.
Both the H-R diagram position and
the association of HD~72106A with the HAeBe 
secondary provide evidence that the primary is 
a pre-main sequence or very early zero-age main 
sequence star.  The H-R diagram of both stars is shown in Fig. \ref{h-r}.

\section{Spectrum Reconstruction}

Ten Stokes $V$ spectra of HD~72106 were obtained between February 2005 and February 2006 
with ESPaDOnS, the high-resolution echelle spectropolarimeter mounted at CFHT. The spectra  
cover a wavelength range from 3700~\AA to just past 1~$\mu$m.
Problematically, the ESPaDOnS pinhole is 1.6'' in diameter, whereas the two 
components of HD~72106 are separated by 0.8''.  With good seeing and careful guiding 
it is possible to isolate individual components. This was accomplished on one occasion. 
However, the majority of our spectra are of combined light from the system. Observations were 
conducted using atmospheric dispersion correction, with careful attention to keeping the system
centroid centred in the pinhole.  

In order to isolate the spectrum of the primary star, we subtracted the individual spectrum of the secondary
(obtained on a night of excellent seeing) 
from the combined spectrum, weighted by its luminosity contribution.  Assuming the stars are located at the same 
distance, their apparent magnitudes, suitably bolometric-corrected, yield their relative luminosities.  
The relative apparent magnitudes of the components of HD~72106 are well known (eg. Fabricius \& Makarov 2000), 
and the Hipparcos parallax solution puts both components at the same distance.  
The subtraction process also assumes that the spectrum of the secondary is non-variable, 
which is a reasonable assumption for metallic lines of a non-magnetic late A star (excluding H$\alpha$).
Thus, with the relative apparent magnitude, we can perform the subtraction effectively.  

\begin{figure}[!htb]
\centering
\includegraphics[width=5.0in]{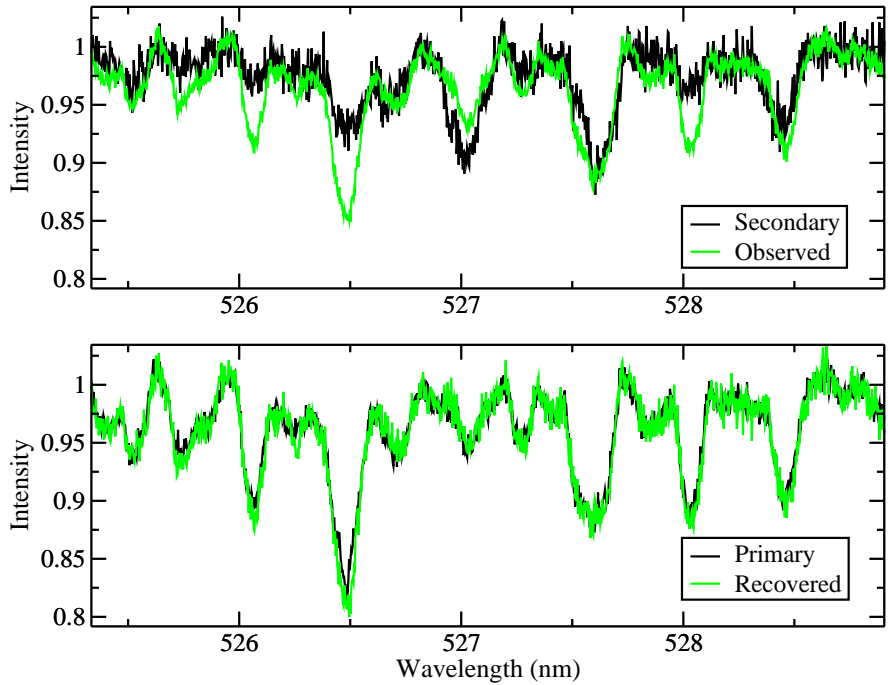}
\caption{The top frame shows a sample of the spectrum 
from the secondary (in black) and an observed spectrum of the combined 
light from the system (in gray).  The bottom frame compares the spectrum reconstructed 
spectrum of the primary (in gray), resulting from the weighted subtraction process, 
with an observation of just the primary (in black).  A very good correspondence is obtained 
(although some differences are expected due to the line profile variability of the primary).  }
\label{sub}
\end{figure}

To check this method, Balmer lines (excluding H$\alpha$), as well as several deep, apparently non-variable metallic lines,  
were compared in the recovered spectra and in the individual observed spectrum of the primary.  
A very good agreement was found.  Presented in Fig. \ref{sub} is a typical observed spectrum of 
the combined system, compared with a spectrum of just the secondary, and the reconstructed 
spectrum of the primary, resulting from the subtraction process, compared with a spectrum 
of just the primary.

\section{Magnetic Field Measurements}

Least Squares Deconvolution (LSD) (Donati et al. 1997) was performed on the reconstructed 
spectra of the primary.  LSD is a method of effectively averaging over 
many lines in a spectrum in order to improve the signal to noise ratio of the line profile.  
A line mask for a 10000 K Bp star, and spectra with the contribution of the 
secondary removed, were used. Significant Stokes V signatures are detected at all observed
phases. 
Longitudinal magnetic field measurements were obtained from the LSD Stokes V 
profile for each observation.  The largest longitudinal field found was $+391 \pm 65$ G.  
Stokes V profiles with larger amplitudes were observed, but with crossover signatures, 
corresponding to weaker longitudinal fields.  The Stokes I and V LSD profiles 
corresponding to the 391 G measurement are shown in Fig. \ref{ivp}, 
and are labeled 0.4015 in the phase column.

\section{Period Searching}

Clear variability, presumably due to rotation, can be seen in the spectrum of HD 72106A, 
with an apparent period around two days.  To investigate this further we first searched 
our measured longitudinal magnetic field data for a period.  
This was done using a modified Lomb-Scargle technique, fitting sinusoids through the data 
to produce a periodogram, and looking for a minimum in $\chi^2$.  
Unfortunately, due to the small number of longitudinal 
field observations, a unique solution could not be found.

In an attempt to improve our period determination, a more sophisticated technique was used.  
This method is still based on fitting a sinusoid through a time series of data points to 
produce a periodogram.  However, now the set of measurements for one pixel in the LSD profile 
is used as the time series of data points.  The periodogram for one data point may be heavily affected 
by noise, but if one averages the periodograms of all the pixels in the LSD line profile, producing an average periodogram, 
the effects of noise are substantially reduced and the true period found.  To verify the technique, periods were 
determined for a number of known Ap stars (Auriere et al., in preparation), as well as 
synthetic test spectra, with good results.  

\begin{figure}[!htb]
\centering
\includegraphics[width=4.0in]{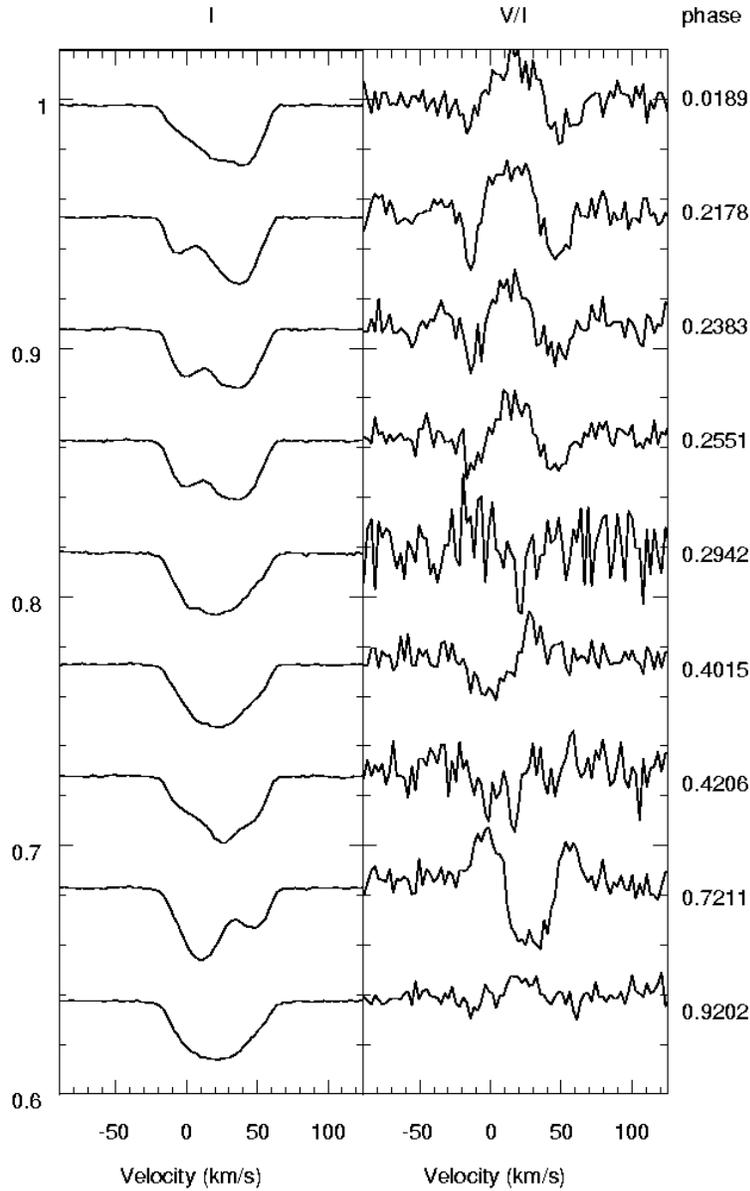}
\caption{The Stokes I and V profiles obtained for HD~72106A, phased according 
to a 1.953 day period. The Stokes V profiles, detected significantly at all phases, have been multiplied by 20 times relative 
to the I profiles, and shifted vertically for clarity.  The profiles are labeled by phase, 
with an arbitrary zero point. Clear variations of the Stokes I profile are apparent.}
\label{ivp}
\end{figure}

While this improved the situation significantly by reducing the 
number of possible periods, it still did not produce a unique period
solution for HD 72106A.  Examining phased LSD profiles by eye allowed 
us to reject several more periods as un-physical, but a few possibilities still remain 
around 1.8 and 1.9 days.  A set of LSD profiles phased with one of the 
better candidate periods is presented in Fig. \ref{ivp}.  

\section{Spectrum Synthesis}

Preliminary spectrum modeling has been performed for both the primary and the secondary, 
for a single phase only, using observations of the individual components.  The Zeeman2 spectrum 
synthesis code was used, with ATLAS9 model atmospheres corresponding to the temperatures and 
gravities deduced by Wade et al. (2006).  In the primary, a dipole magnetic field of 1 kG was assumed, 
oriented along the line of sight, with no microturbulence.  A homogeneous abundance distribution both 
vertically and horizontally was assumed.  In the secondary, a homogeneous vertical and horizontal 
abundance distribution was used, with no magnetic field.  Microturbulence was left as a free 
parameter in the fitting procedure.  

We find that the primary displays clear chemical 
peculiarities. The abundance pattern is typical of those of Ap/Bp stars, although the abundance 
enhancements are remarkably strong.  In particular, strong overabundances of 
Cr, Fe, Si, and Ti are found.  In contrast, the secondary displays solar abundance
for all elements detected.  

Clear structure and variability can be seen in the line profile of the primary, 
due to abundance spots on the surface of the star.  These appear most clearly in lines of Fe and Cr, 
but can also be observed in some Ti and Si lines.  
Complex, variable line profiles of this type are typical of Ap stars, and are explained by the presence of abundance non-uniformities in the stellar photosphere.
The secondary displays smooth line profiles with no visible asymmetries (although the S/N of this star is relatively poor, and additional observations are required to confirm this).

\begin{figure}[!htb]
\centering
\includegraphics[width=4.0in]{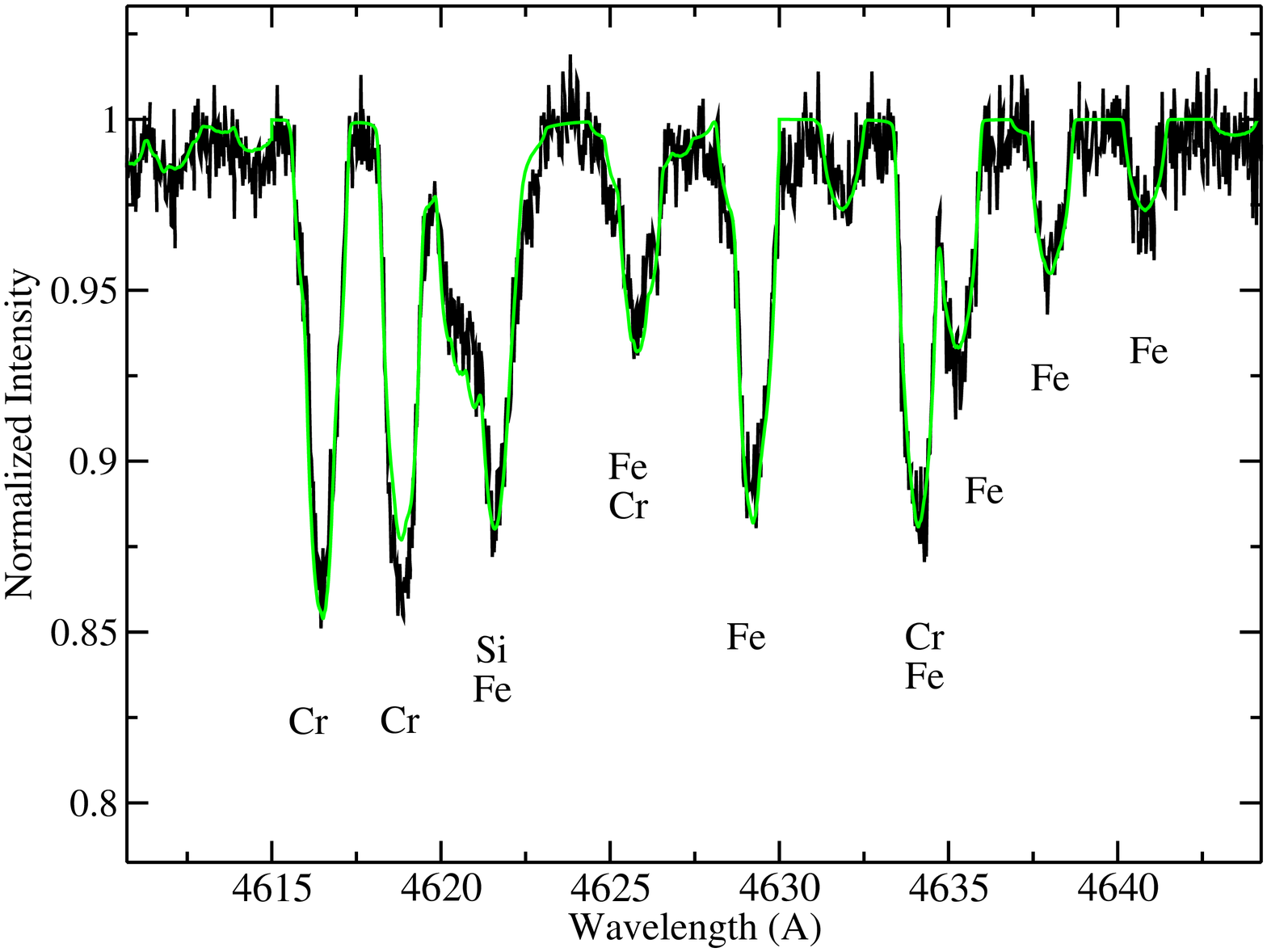}
\caption{Observed (black) and best fit synthetic (gray) spectra for HD~72106A.  
Major contributers to each feature have been labeled. }
\label{f-abun-A}
\end{figure}

\begin{figure}[!htb]
\centering
\includegraphics[width=4.0in]{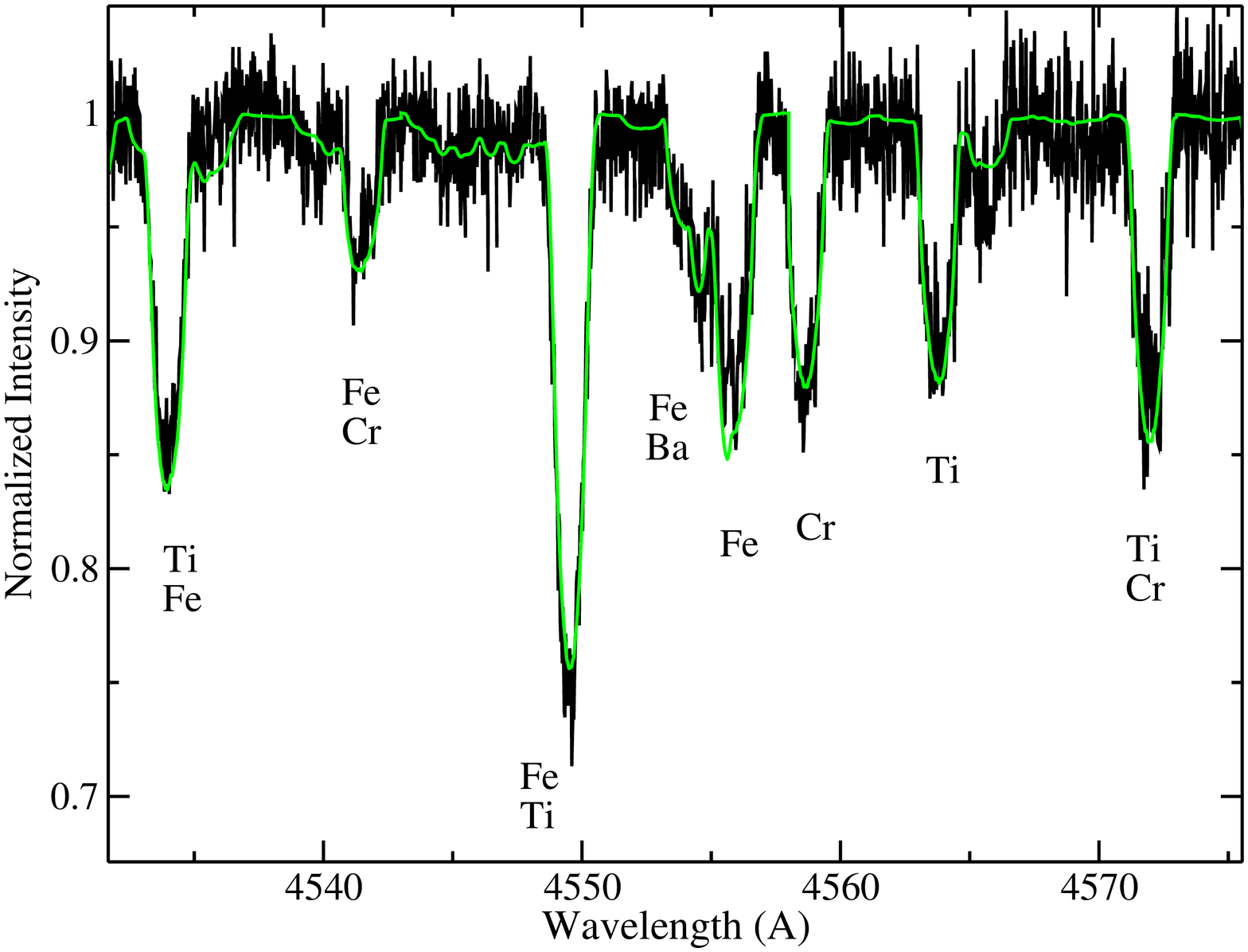}
\caption{Observed (black) and best fit synthetic (gray) spectra for HD~72106B.  
Major contributers to each feature have been labeled. }
\label{f-abun-B}
\end{figure}

A sample spectrum of the primary, with a best fit synthetic spectrum, is shown in Fig. \ref{f-abun-A}.  
Best fit abundances and $v \sin i$ for the primary are shown in Table \ref{t-abun}.
For the secondary, best fit abundances, $v \sin i$, and microturbulence 
can also be seen in Table \ref{t-abun}, with sample observed and synthetic spectra in Fig. \ref{f-abun-B} .

\begin{table}[!htb]
\begin{center}

\begin{tabular}{cccc}
\hline \hline \noalign{\smallskip}
Element  & Primary &  Secondary & \\
\noalign{\smallskip} \hline \noalign{\smallskip}
Al & $-0.2 \pm 0.4$ & \\
Si & $+1.0 \pm 0.2$ & \\
Ca &                & $-0.4 \pm 0.4$ \\
Ti & $+1.0 \pm 0.3$ & $-0.1 \pm 0.4$ \\
Cr & $+1.9 \pm 0.1$ & $+0.1 \pm 0.3$ \\
Fe & $+0.9 \pm 0.1$ & $-0.1 \pm 0.3$ \\
Ba &                & $-0.3 \pm 0.5$ \\
\noalign{\smallskip} \hline \noalign{\smallskip}
$v \sin i$ & $41 \pm 3$ km\,s$^{-1}$ & $50 \pm 4$ km\,s$^{-1}$ \\
$\xi$      & $\sim 0$ km\,s$^{-1}$   & $2.5 \pm 0.5$ km\,s$^{-1}$ \\
\noalign{\smallskip} \hline \hline

\end{tabular}
\end{center}
\caption{Preliminary best-fit abundances, $v \sin i$, and microturbulence values for HD~72106A and HD~72106B.  
Abundances are based on the 4500-4700\AA\, region.  }
\label{t-abun}

\end{table}

\section{Conclusions}

These results demonstrate that strong chemical peculiarities, of the type seen 
in Ap/Bp stars, together with magnetic fields, can exist in late B-type stars at 
the pre-main sequence or very early zero-age main sequence phase.  
A maximum longitudinal magnetic field of  $+391 \pm 65$ G was found in HD~72106A. 
Overabundances of up to 2 dex above solar were also found, along with a 
complex absorption line structure, suggesting an inhomogeneous surface 
abundance distribution of elements.  These properties are all characteristic of Ap stars.  
In contrast, HD~72106B which presumably formed at the same point in time, 
under the same conditions as the primary, displays solar abundances and no magnetic field.  
Thus the HD~72106 system provides a unique link between Ap stars and their pre-main sequence progenitors.  

Further circular polarisation observations of the HD~72106 system are currently planned. 
This will allow us to better sample the spectroscopic and polarimetric cycle of the primary to determine a 
unique rotational period.  With additional observations we can model the magnetic field geometry 
and map the surface distributions of elements.  Additional observations will also allow us to
test the stability of the spectrum of the secondary and further test the spectrum reconstruction 
technique.  With these new observations we will be able to more completely describe this unique and remarkable system.

\end{document}